\documentclass[3p,times,twocolumn,compress]{elsarticle}
\usepackage{ecrc}

\pdfoutput=1

\volume{00}
\firstpage{1}
\journalname{Nuclear Physics B (Proc. Suppl.)}
\runauth{A.V. Gramolin et al.}
\jid{nuphbp}
\jnltitlelogo{Nuclear Physics B Proceedings Supplements}

\usepackage{amsmath,amssymb}
\usepackage{hyperref}
\usepackage{color}

\begin{document}

\begin{frontmatter}
\title{Measurement of the two-photon exchange contribution\\ in elastic $ep$~scattering at \mbox{VEPP--3}}

\author[BINP]{A.V.~Gramolin\corref{cor}}
\ead{gramolin@inp.nsk.su}
\ead[url]{www.inp.nsk.su/~gramolin/}

\author[ANL]{J.~Arrington}
\author[BINP]{L.M.~Barkov}
\author[BINP,NSU]{V.F.~Dmitriev}
\author[TPU]{V.V.~Gauzshtein}
\author[BINP]{R.A.~Golovin}
\author[ANL]{R.J.~Holt}
\author[BINP]{V.V.~Kaminsky}
\author[BINP]{B.A.~Lazarenko}
\author[BINP]{S.I.~Mishnev}
\author[BINP,NSU]{N.Yu.~Muchnoi}
\author[BINP]{V.V.~Neufeld}
\author[BINP]{D.M.~Nikolenko}
\author[BINP]{I.A.~Rachek}
\author[BINP]{R.Sh.~Sadykov}
\author[BINP,NSU]{Yu.V.~Shestakov}
\author[TPU]{V.N.~Stibunov}
\author[BINP,NSU]{D.K.~Toporkov}
\author[NIKHEF]{H.~de~Vries}
\author[BINP]{S.A.~Zevakov}
\author[BINP]{V.N.~Zhilich}

\cortext[cor]{Corresponding author}

\address[BINP]{Budker Institute of Nuclear Physics of SB~RAS, Novosibirsk, Russia}
\address[ANL]{Argonne National Laboratory, Argonne, USA}
\address[NSU]{Novosibirsk State University, Novosibirsk, Russia}
\address[TPU]{Nuclear Physics Institute of Tomsk Polytechnic University, Tomsk, Russia}
\address[NIKHEF]{NIKHEF, Amsterdam, The Netherlands}

\begin{abstract}
We report on the status of the Novosibirsk experiment on a precision measurement of the ratio $R$ of the elastic $e^+ p$ and $e^- p$ scattering cross sections. Such measurements determine the two-photon exchange effect in elastic electron-proton scattering. The experiment is conducted at the \mbox{VEPP--3} storage ring using a~hydrogen internal gas target. The ratio~$R$ is measured with a beam energy of~1.6~GeV (electron/positron scattering angles are $\theta = 55{\div}75^{\circ}$ and $\theta = 15{\div}25^{\circ}$) and 1~GeV ($\theta = 65{\div}105^{\circ}$). We briefly describe the experimental method, paying special attention to the radiative corrections. Some preliminary results are presented.
\end{abstract}

\end{frontmatter}

\section{Introduction}

The elastic electromagnetic form factors of the proton are important characteristics that describe its internal structure. More precisely, the electric $G_E \bigl(Q^2\bigr)$ and magnetic $G_M \bigl(Q^2\bigr)$ form factors describe the spatial distributions of the proton's charge and magnetization as functions of four-momentum transfer squared~$Q^2$ (see Refs.~\cite{Hyde-Wright(2004), Perdrisat(2007)} for reviews). Until recently these form factors were determined only by an analysis of differential cross section of the unpolarized elastic $ep$~scattering.

\begin{figure}[t]
\centering
\includegraphics[width=0.14\textwidth]{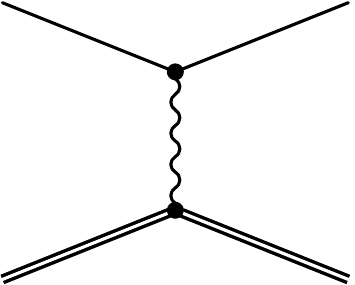} \hspace{0.5mm} \includegraphics[width=0.14\textwidth]{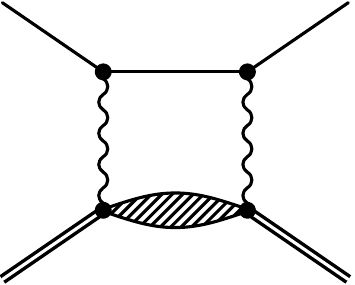} \hspace{0.5mm} \includegraphics[width=0.14\textwidth]{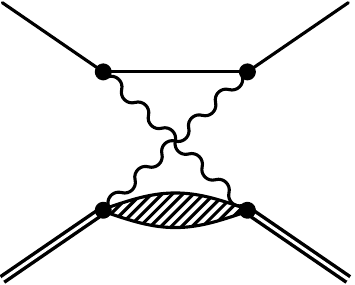}
\caption{Feynman diagrams of the two lowest orders in~$\alpha$ for the elastic electron-proton scattering. The first diagram corresponds to the one-photon exchange approximation. The box and crossed-box diagrams describe the two-photon exchange (the blob represents possible intermediate states of the proton).}
\label{Fig1}
\end{figure}

Let us recall that the differential cross section $d \sigma / d \Omega$ for elastic $ep$~scattering in the one-photon exchange approximation (see the first diagram in Fig.~\ref{Fig1}) is given by the Rosenbluth formula~\cite{Rosenbluth(1950)}:
$$
\frac{d \sigma}{d \Omega} = \frac{\tau}{\varepsilon \, (1 + \tau)} \left[G_M^2 \bigl(Q^2\bigr) + \frac{\varepsilon}{\tau} G_E^2 \bigl(Q^2\bigr)\right] \frac{d \sigma_\text{Mott}}{d \Omega},
$$
where $\tau = Q^2 / \bigl(4 M^2\bigr)$, $\varepsilon = \bigl[1 + 2 \, (1 + \tau) \tan^2{(\theta / 2)}\bigr]^{-1}$ is the virtual photon polarization, $\theta$~is the electron scattering angle in the laboratory frame and $d \sigma_{\text{Mott}} / d \Omega$~is the Mott differential cross section. It follows from this formula that the proton form factors can be determined by measuring the differential cross section at fixed momentum transfer~$Q^2$, but with different electron scattering angles and incident beam energies.

In the mid-nineties, it became possible to use another method to study the proton form factors, the polarization transfer method, which was originally proposed back in~1974~\cite{Akhiezer&Rekalo(1974), Arnold(1981)}. In this method a polarization of the recoil proton in the process of elastic scattering of longitudinally polarized electrons on an unpolarized hydrogen target is measured. In such a case the ratio $G_E / G_M$ is directly proportional to the ratio of transverse and longitudinal polarizations of recoil protons.

\begin{figure}[t]
\centering
\includegraphics[width=0.47\textwidth]{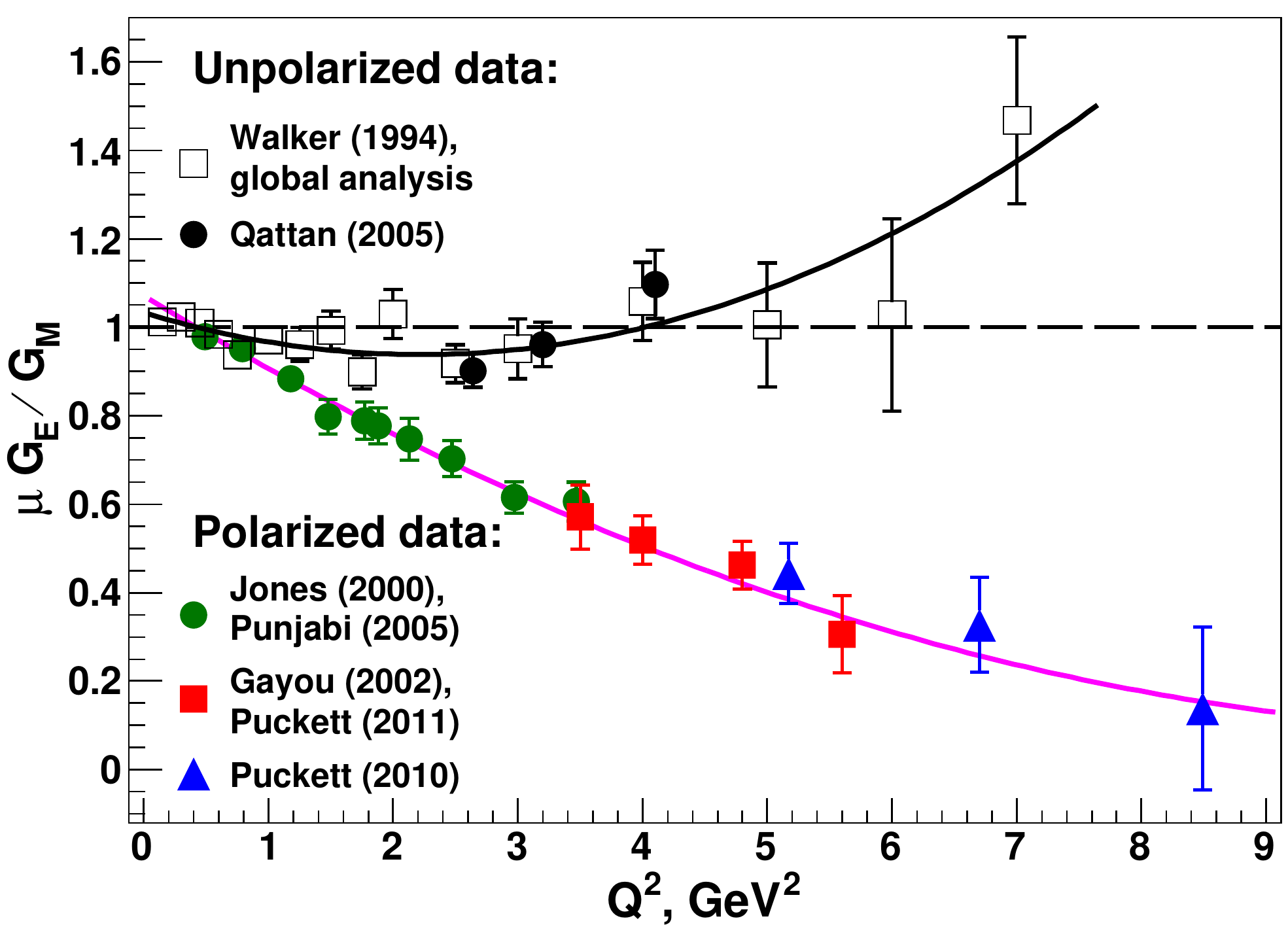}
\caption{Comparison of the polarized~\cite{Jones(2000), Punjabi(2005), Gayou(2002), Puckett(2011), Puckett(2010)} and unpolarized~\cite{Walker(1994), Qattan(2005)} data for the ratio $\mu \, G_E / G_M$ (where $\mu$ is the proton magnetic moment). The solid lines are the quadratic fits for these two groups of data.}
\label{Fig2}
\end{figure}

Measurements performed at TJNAF with the polarization transfer method~\cite{Jones(2000), Punjabi(2005), Gayou(2002), Puckett(2011), Puckett(2010)} gave the results that are in contradiction with the results obtained by the Rosenbluth method. This is clearly seen in Fig.~\ref{Fig2}, where the polarized data of TJNAF as well as several results of the Rosenbluth measurements are shown. The latter include the global analysis by Walker~\cite{Walker(1994)} and new accurate Rosenbluth measurement~\cite{Qattan(2005)}.

The most probable cause of this discrepancy is the failure of the one-photon exchange approximation to precisely describe the results of unpolarized measurements. However, application of the two-photon exchange corrections (see the diagrams in Fig.~\ref{Fig1}) runs into difficulties: on the one hand, there are no reliable calculations and, on the other hand, there are no sufficiently accurate experimental data (for review see Ref.~\cite{Arrington(2011)}).

The two-photon exchange effect can be measured experimentally by precisely comparing the $e^-p$ and $e^+ p$ scattering~\cite{Arrington(2011)}. It is possible because the interference between the one-photon exchange amplitude and the two-photon exchange amplitude occurs with opposite sign for electrons and positrons. In 2004, it was proposed~\cite{VEPPproposal(2004)} to perform such measurement at the \mbox{VEPP--3} storage ring~\cite{VEPP-3} in Novosibirsk. Here we report on the status of this experiment. It should be mentioned that there are also two similar experiments underway at DESY (OLYMPUS collaboration~\cite{OLYMPUS}) and at TJNAF (CLAS collaboration~\cite{CLAS}).

\section{Description of the experiment}

The experiment was divided into two runs with different kinematic parameters (for details see Table~\ref{Tab}). The first run (with a beam energy of 1.6~GeV) was conducted in late 2009. Analysis of data obtained in this run is continuing. The preliminary results are presented here (also see Ref.~\cite{Nikolenko(2010)}). The second run (with a beam energy of~1~GeV) is underway and will continue, presumably, until March 2012.

\begin{table}[b]
\centering
\begin{tabular}{|c|c|c|c|c|c|}
\hline & \multicolumn{3}{|c|}{Run I} & \multicolumn{2}{|c|}{Run II} \\
\cline{2-6} Parameter & LA & MA & SA & LA & MA \\
\hline \hline \rule{0mm}{3.6mm} $E_{\text{beam}}$, GeV & \multicolumn{3}{|c|}{1.6} & \multicolumn{2}{|c|}{1.0} \\
\hline \rule{0mm}{3.6mm} $\theta$, degree & 55/ & 15/ & 8/ & 65/ & 15/ \\
(min/max) & 75 & 25 & 15 & 105 & 25 \\
\hline \rule{0mm}{3.6mm} $Q^2, \text{ GeV}^2$ & 1.26/ & 0.16/ & 0.05/ & 0.71/ & 0.07/ \\
(min/max) & 1.68 & 0.41 & 0.16 & 1.08 & 0.17 \\
\hline \rule{0mm}{3.6mm} $\varepsilon$ & 0.58/ & 0.97/ & 0.99/ & 0.51/ & 0.97/ \\
(max/min) & 0.37 & 0.90 & 0.97 & 0.18 & 0.91 \\
\hline
\end{tabular}
\caption{Kinematic parameters of the experiment.}
\label{Tab}
\end{table}

The ratio $R = \sigma \bigl(e^+ p\bigr) / \sigma \bigl(e^- p\bigr)$ of the elastic positron-proton and electron-proton scattering cross sections is measured. The scattered electron/positron and the recoil proton are detected in coincidence. In the first run, the particles were detected in three angular ranges: large angles (LA), medium angles (MA) and small angles (SA). Scattering of $e^- / e^+$ at small angles was used only for luminosity monitoring (this is possible because the two-photon exchange effect is negligible for this kinematic range). In the second run, the particles are detected in two angular ranges: large angles (LA) and medium angles (MA). The latter range is used for luminosity monitoring. Kinematic parameters for both runs are shown in Table~\ref{Tab}.

The hydrogen internal gas target is used, similar to that previously used in the experiments at \mbox{VEPP--3}~\cite{Dyug(2002)}. The target is an open-ended storage cell, which has an elliptical cross-section $13 \times 24 \text{ mm}^2$ and a length of 400~mm. High-purity hydrogen gas is directed into the center of the storage cell. The thickness of the target is about~$10^{15} \text{ at.}/\text{cm}^2$. Hydrogen is pumped from the \mbox{VEPP--3} vacuum chamber using cryogenic pumps.

\begin{figure}[t]
\centering
\includegraphics[width=0.47\textwidth]{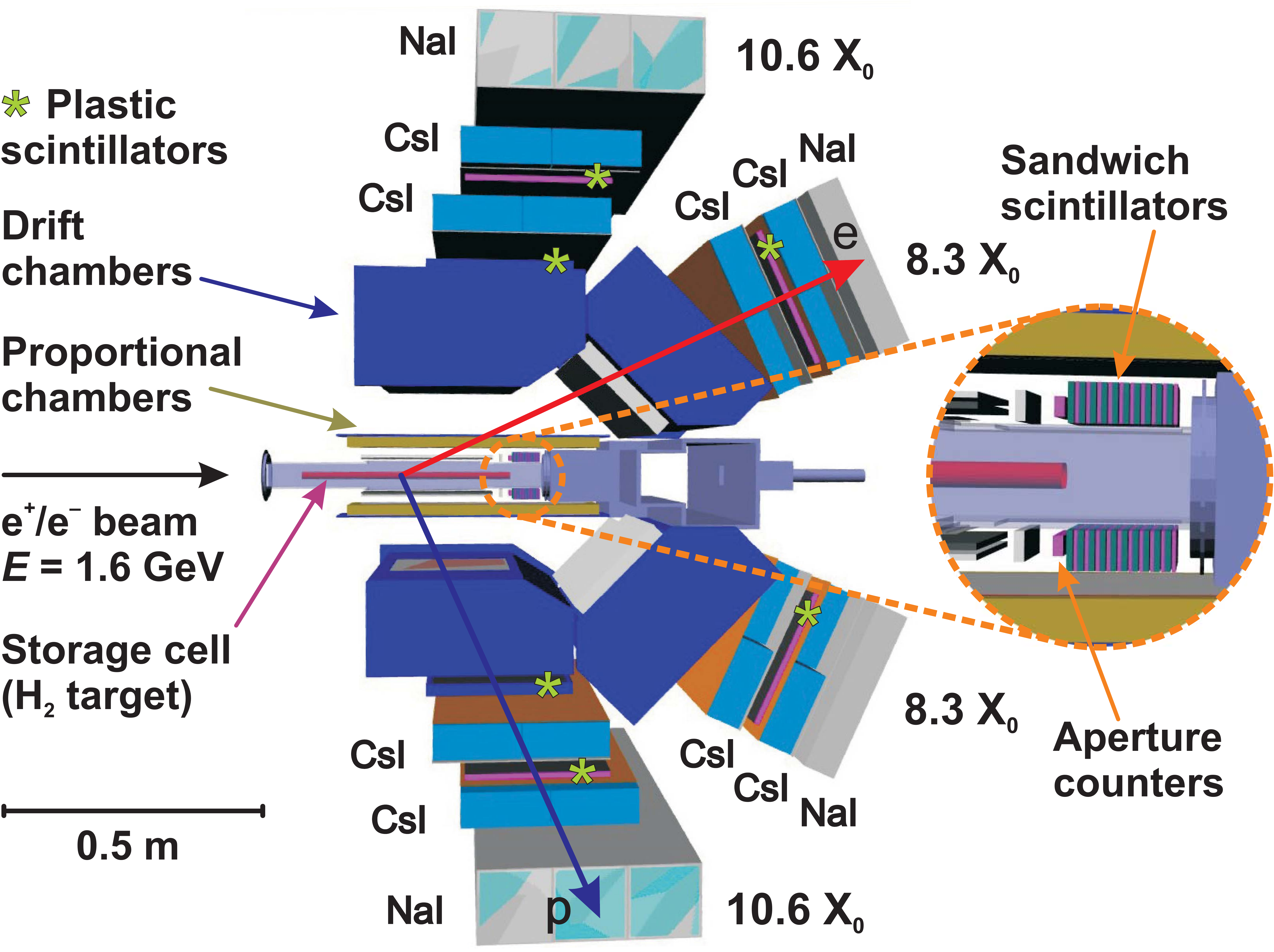}
\caption{Schematic side view of the detector used in the first run.}
\label{Fig3}
\end{figure}

The detector (see Fig.~\ref{Fig3}) consists of two identical parts (top and bottom) arranged symmetrically with respect to the median plane of the storage ring. Azimuthal angle acceptance of the detector is~$\Delta \phi = 2 \cdot 60^{\circ} = 120^{\circ}$. In the first run, the particles were detected in three angular ranges (LA, MA and SA). In the LA and MA ranges the detector has a tracking system (drift and multi-wire proportional chambers), calorimeters (CsI and NaI crystals) and plastic scintillators used for event triggering and for time-of-flight measurements. As for the SA range, here we used the sandwich scintillators.

In the second run, we have a slightly different configuration of the detector, but it is similar to that described above. An important feature is that here we have the possibility to reconstruct the full kinematics for some of the events with emission of hard real photons. This will allow us to test our calculations for the first-order bremsstrahlung (see the next section).

We use the following cuts (in various combinations for different ranges) for the selection of elastic scattering events: correlation between the polar angles of~$e^-$/$e^+$ and~$p$; the same for the azimuthal angles; correlation between the lepton scattering angle and the proton energy; correlation between the lepton scattering angle and the lepton energy; $\Delta E$--$E$ analysis for the proton identification; and time-of-flight analysis for the identification of low-energy protons. The main sources of background events are pion production processes and machine backgrounds.

To suppress systematic errors it is important to maintain the same experimental conditions during data taking with electron and positron beams. Therefore, data collection with $e^-$ and~$e^+$ beams is alternated regularly. This allows us to suppress effects of slow drift in time of the target thickness, detection efficiency and some other parameters.

A typical working cycle of the experiment consists of the following stages: storing of $e^-$/$e^+$ beam up to $40 \div 55$~mA; ramping beam energy (from the injection energy of 0.35~GeV to~1.6 or 1~GeV); switching on the internal gas target and then data taking (as long as the beam current drops to $10 \div 13$~mA); and returning to the injection energy. The full cycle with two beams (electron and positron) takes $60 \div 80$ minutes, from which the data taking occupies approximately 40~minutes. The average luminosity of the experiment is about $5 \cdot 10^{31} \text{ s}^{-1} \cdot \text{cm}^{-2}$. The beam current integral collected in the first run is~54~kC. In the second run, we plan to collect the beam current integral of $60 \div 90$~kC.

The main possible sources of systematic errors of the measurement are as follows: different beam positions for electrons and positrons; unequal beam energy for electrons and positrons; time instability of detection efficiency; drift of the target thickness in time; and uncertainty related to the radiative corrections. For the first run, the total systematic uncertainty is now estimated as~$3 \cdot 10^{-3}$ (for both LA and MA ranges).

There are three sources of information on the beam position: \mbox{VEPP--3} beam position monitors (electrostatic pick-ups); movable beam scrapers; and reconstruction of the event vertex position using the tracking system of the detector. As for the beam energy, it is measured regularly and with good accuracy (about 50~keV) during the data taking by applying the Compton backscattering technique~\cite{Blinov(2009), Blinov(ICFA)}. The essence of this technique is as follows. Photons from a~$\text{CO}_2$ laser are scattered in a head-on collision with the stored beam. The beam energy can be determined from the spectrum of the backscattered photons that are detected by a high-purity germanium detector.

\section{Radiative corrections}

\begin{figure}[t]
\centering
\includegraphics[width=0.112\textwidth]{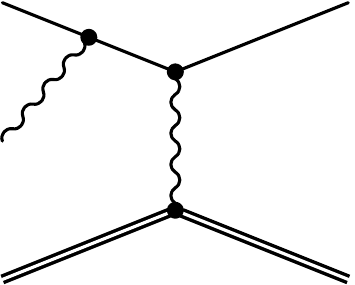} \includegraphics[width=0.112\textwidth]{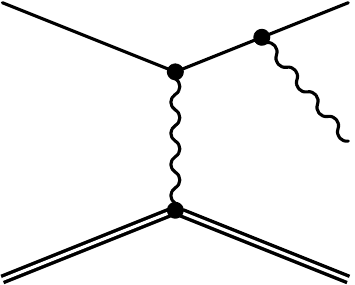} \includegraphics[width=0.112\textwidth]{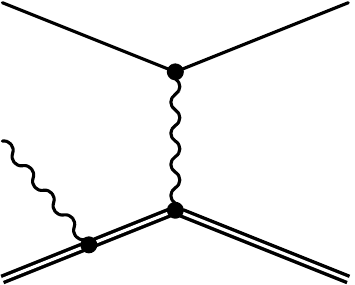} \includegraphics[width=0.112\textwidth]{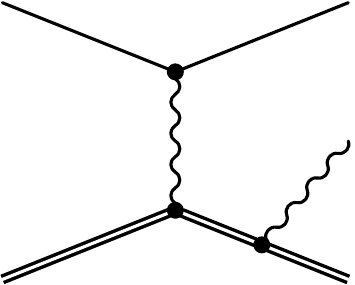}
\caption{Feynman diagrams representing the first-order internal bremsstrahlung.}
\label{Fig4}
\end{figure}

\begin{figure}[t]
\centering
\includegraphics[width=0.14\textwidth]{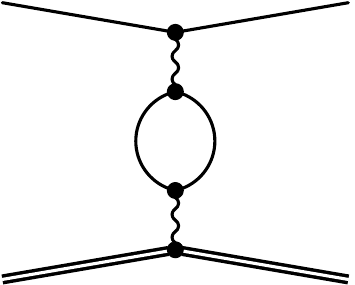} \hspace{0.5mm} \includegraphics[width=0.14\textwidth]{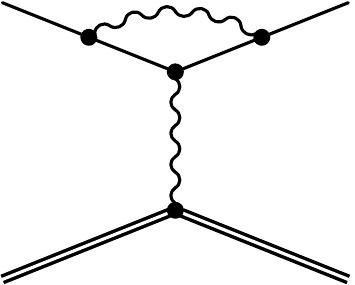} \hspace{0.5mm} \includegraphics[width=0.14\textwidth]{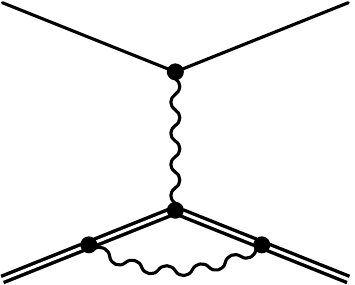}
\caption{Feynman diagrams representing virtual photon corrections.}
\label{Fig5}
\end{figure}

Not only the two-photon exchange, but also other higher-order QED processes affect the elastic scattering cross section. These processes include internal bremsstrahlung of real photons (see the diagrams in Fig.~\ref{Fig4}) and the processes associated with virtual photons (vacuum polarization and vertex corrections, see Fig.~\ref{Fig5}). Note that the external bremsstrahlung is neglected due to the small thickness of the target.

An important fact is that the internal bremsstrahlung affects the measured ratio~$R$ (because the interference between the lepton bremsstrahlung and the proton bremsstrahlung occurs with opposite sign for electrons and positrons). The bremsstrahlung process can easily lead to a difference of a few percent between the cross sections of $e^- p$ and $e^+ p$ elastic scattering. It is therefore very important to carefully consider the radiative corrections due to bremsstrahlung.

In addition, the radiative corrections due to brems\-strahlung depend on the detector geometry, its spatial and energy resolutions and the kinematic cuts used in event selection. Accounting for all these factors is only possible with careful simulation of the detector. For this purpose, a new Monte~Carlo event generator~\cite{ESEPP}, called \mbox{ESEPP} (Elastic Scattering of Electrons and Positrons by Protons), has been developed. The generated events ($e^{\pm} p \rightarrow e^{\pm} p$ and $e^{\pm} p \rightarrow e^{\pm} p \, \gamma$) are used for Geant4 simulation of the detector response.

We apply the exact calculation~\cite{Fadin&Feldman} for the first-order bremsstrahlung. The main features of this calculation are as follows. First, we do not use the peaking approximation or even the soft photon approximation. Second, we consider both lepton bremsstrahlung and proton bremsstrahlung. Third, in the proton line we use the full electromagnetic vertex operator~$\Gamma^{\mu}$ (see Ref.~\cite{Ent(2001)}) for the real photon emission. In addition, we apply the calculation~\cite{Fadin&Gerasimov} for the first-order bremsstrahlung with the delta-isobar $\Delta (1232)$ excitation. We use this calculation as an estimate of the contribution of the first-order bremsstrahlung with the proton excitations in the intermediate state. A detailed description of these calculations and the ESEPP event generator will be published elsewhere.

As for the other processes (vacuum polarization and vertex corrections), we use the standard approach described in Refs.~\cite{Ent(2001), Mo&Tsai}. Multiple soft photon emission~\cite{Ent(2001)} is not taken into account, as the estimates show that it can be neglected under our experimental conditions. As usual, only the infrared divergent parts of the two-photon exchange amplitude and the proton vertex correction are taken into account. These terms exactly cancel with the corresponding infrared divergent terms from the bremsstrahlung process.

\section{Preliminary results}

\begin{figure}[t]
\centering
\includegraphics[width=0.47\textwidth]{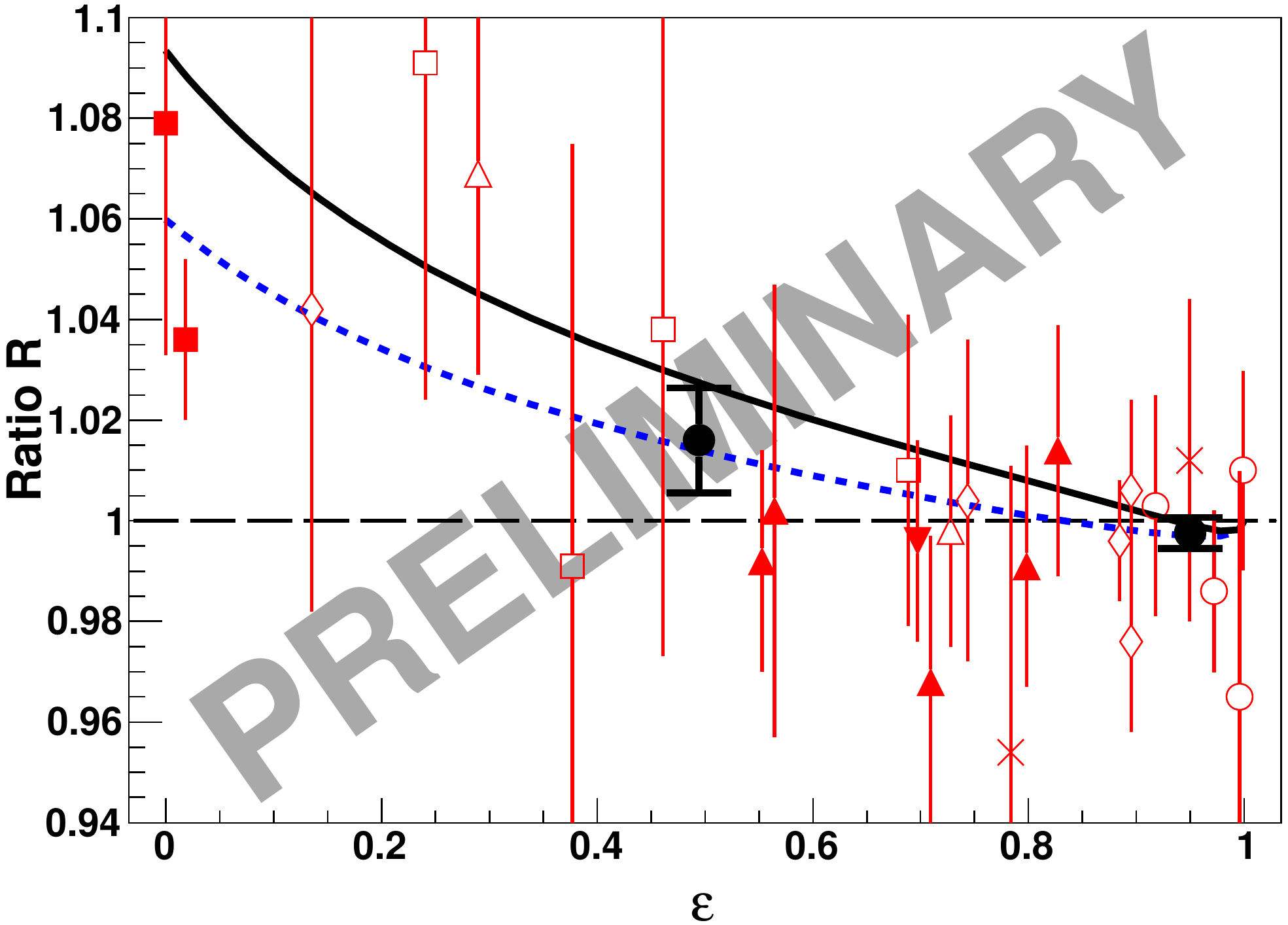}
\caption{Comparison of the preliminary results for the ratio~$R$ (radiatively corrected) with the previous measurements (for $Q^2 < 2 \; \text{GeV}^2$): \textcolor{red}{\normalsize $\lozenge$}~\cite{Yount(1962)}; \textcolor{red}{$\square$}~\cite{Browman(1965)}, first experiment; \textcolor{red}{\large $\triangle$}~\cite{Browman(1965)}, second experiment; \textcolor{red}{\large $\blacktriangledown$}~\cite{Anderson(1966)}; \textcolor{red}{\large $\times$}~\cite{Bartel(1967)}; \textcolor{red}{$\blacksquare$}~\cite{Bouquet(1968)}; \textcolor{red}{\large $\blacktriangle$}~\cite{Anderson(1968)}; \textcolor{red}{\Large $\circ$}~\cite{Mar(1968)}; {\Large \textbullet}~this experiment, run~I. The curves show the theoretical predictions~\cite{Blunden(2005), Blunden} for the ratio~$R$ due to the two-photon exchange. The solid (dashed) line corresponds to the kinematics of run~I (run~II).}
\label{Fig6}
\end{figure}

Preliminary results of the first run are shown in Fig.~\ref{Fig6} in comparison with all available world data on the ratio $R$ (in the range $Q^2 < 2 \text{ GeV}^2$). It can be seen that our results have better accuracy than the data available. For the LA range ($\epsilon = 0.5$, $Q^2 = 1.43 \text{ GeV}^2$) the result is $R = 1.016 \pm 0.011 \pm 0.003$, and for the MA range ($\epsilon = 0.95$, $Q^2 = 0.23 \text{ GeV}^2$) the result is $R = 0.9976 \pm 0.0009 \pm 0.003$. Here, the first error is statistical and the second is systematic. These results are preliminary because some minor corrections have not yet been made (for example, corrections related to the variation of beam energy and position). Note that before applying the radiative corrections, these values were $R = 1.053$ for the LA range and $R = 1.0044$ for the MA range.

The curves in Fig.~\ref{Fig6} show the theoretical predictions for the ratio~$R$ due to the two-photon exchange~\cite{Blunden(2005)}. These data were kindly provided by Prof.~Blunden~\cite{Blunden}. The solid and dashed lines correspond to the kinematics of the first and the second runs, respectively. If these predictions are correct, then the expected size of the two-photon exchange effect in the second run should be approximately the same as for the LA range in the first run. At the same time, the elastic $e^{\pm} p$ scattering cross sections are larger for the kinematics of the second run.

The second run of the experiment is underway, and we are not ready to present results yet. We plan to get results in two or three experimental points in the range $\varepsilon = 0.2 \div 0.5$ with total errors of about~$6 \cdot 10^{-3}$.

\section{Conclusion}

We have described the status of the Novosibirsk experiment on the precise comparison of the $e^- p$ and $e^+ p$ elastic scattering cross sections. The results of our measurement are more accurate than the data currently available. The experimental data give the size of the two-photon exchange effect in elastic electron-proton scattering. This information is important to explain the recently observed discrepancy between new polarized measurements and old data on the proton electromagnetic form factors.

Special attention is given to the radiative corrections, in particular, to the first-order bremsstrahlung. We have developed the event generator, which is based on the accurate calculation of the first-order bremsstrahlung beyond the soft-photon approximation. The generated events are used for Geant4 simulation of the detector response. This allows us to take into account the geometry of the detector, its coordinate and energy resolutions and kinematic cuts used in event selection.

We have presented the preliminary results of the first run. The second run of the experiment is underway. It is too early to draw final conclusions based on the results presented, since the data analysis is still ongoing. It can be noted, however, that the preliminary results are consistent with the theoretical predictions~\cite{Blunden(2005), Blunden}.

\section{Acknowledgments}

We acknowledge the staff of \mbox{VEPP--3} for excellent performance of the storage ring during the experiment. We are very grateful to V.S.~Fadin, A.L.~Feldman and R.E.~Gerasimov for their help in calculations of the radiative corrections. This work was supported in part by Russian Foundation for Basic Research under grants 08-02-00624-a, 08-02-01155-a, 10-02-08433-z; by Federal Agency for Education under State Contract P522; by Federal Agency for Science and Innovations under Contract 02.740.11.0245.1; by U.S. Department of Energy under grant DE-AC02-06CH11357; and by U.S. National Science Foundation under grant PHY-03-54871.

\end{document}